\documentstyle[epsf,prb,aps,floats,twocolumn]{revtex}

\begin{document} 
 \draft
\wideabs{ 

\title{
Charge Gap in the One-Dimensional Extended Hubbard Model at Quarter Filling
}

\author{K.  Sano}
\address{Department of Physics Engineering,  Mie University, Tsu, Mie
 514-8507 
, Japan} 
\author{Y. \=Ono}
\address{Department of Physics, Niigata University, Ikarashi, Niigata
 950-2181
, Japan
\\Center for Transdisciplinary Research, Niigata University, Ikarashi, Niigata 950-2181
}

\date{\today}
\maketitle

\begin{abstract}
We propose a new combined approach of the exact diagonalization, the  renormalization  group  and the Bethe ansatz for precise estimates of the charge gap  $\Delta$ in the one-dimensional extended Hubbard model with the onsite and the nearest-neighbor interactions $U$ and $V$ at quarter filling. This approach enables us to obtain the absolute value of $\Delta$ including the prefactor without ambiguity even in the critical regime of the metal-insulator transition (MIT) where $\Delta$ is exponentially small, beyond usual  renormalization  group methods and/or finite size scaling approaches. 
The detailed results of $\Delta$ down to of order of $10^{-10}$ near the MIT are shown as contour lines on the $U$-$V$ plane.

\end{abstract} 
}


There has been much theoretical interest in the one-dimensional (1D)  strongly correlated electron systems such as  the $t$-$J$ model and the  Hubbard  model as  a good testing  ground for the concept of the  Tomonaga-Luttinger liquid \cite{Emery1,Solyom}. Various methods, such as the weak coupling theory (g-ology), the bosonization theory, the Bethe ansatz (BA) method, the conformal  field  theory and the numerical approaches have been  used to clarify  the nature  of these models \cite{Voit}. Among them, combined approaches of the exact diagonalization (ED) and the renormalization group (RG) methods have been extensively developed to investigate the critical behavior near the quantum critical point \cite{Emery-Noguera,Okamoto,Nakamura,Sano1,Sano2,Sano3}. These approaches enable us to obtain accurate results of the phase  boundary for the spin gap phase and those for the charge gap phase beyond the purely numerical approaches combined with the usual finite size scaling.

Recently, we have intensively examined the critical behavior  near the metal-insulator transition (MIT) in the one-dimensional extended Hubbard model with the on-site and the nearest-neighbor interactions $U$ and $V$ at quarter filling using a combined approach of the ED and RG methods \cite{Sano1,Sano2,Sano3}. In this approach, the Luttinger-liquid parameter $K_{\rho}$ is calculated by using the ED for finite size systems and is substituted into the RG equation as an initial condition to obtain $K_{\rho}$ in the infinite size system. The obtained result agrees very well with the available exact result for $U=\infty$ even in the critical regime of the MIT where the characteristic energy becomes exponentially small and the usual finite size scaling is not applicable. When the system approaches the MIT critical point $V\to V_c$ for a fixed $U$, $K_{\rho}$ behaves as $(K_{\rho}-\frac{1}{4})^2=c_K(1-V/V_c)$, where the critical value $V_c$ and the coefficient $c_K$ are functions of $U$  \cite{Sano3}. This approach also yields the critical behavior of the charge gap $\Delta$ in the insulating state near the MIT, where $|\ln \Delta|^{-2}=c_\Delta(V/V_c-1)$ with the coefficient $c_\Delta$ which is a function of $U$. In these studies \cite{Sano1,Sano2,Sano3}, however, the absolute value of $\Delta$ including  prefactor was not explicitly obtained.

In general, it is considered to be difficult for the  RG method or its derivative method to yield  absolute value of  physical quantities including  prefactor.  To overcome this difficulty,  we join  the BA method  to our previous combined approach of the ED and the RG methods in the present study. More explicitly, the BA result in the infinite size system with $U=\infty$ is connected to the ED result in the finite size system with finite $U$ through the analysis of the RG solution. The new combined approach enables us to estimate the absolute value of $\Delta$ including the prefactor without ambiguity in contrast to the previous combined approaches.

The extended Hubbard  model is given by the following Hamiltonian
\begin{eqnarray} 
H&=&-t\sum_{i,\sigma} (c_{i\sigma}^{\dagger} c_{i+1\sigma}+h.c.)
        \nonumber \\
    &+&  U\sum_{i}n_{i\uparrow}n_{i\downarrow}
    +V\sum_{i,\sigma \sigma' }n_{i\sigma}n_{i+1\sigma'}, 
 \label{Model}
\end{eqnarray} 
where $c^{\dagger}_{i\sigma}$  stands for the creation operator for an
 electron with spin $\sigma$  at site $i$ and
 $n_{i\sigma}=c_{i\sigma}^{\dagger}c_{i\sigma}$.
$t$ represents the transfer energy between the nearest-neighbor sites and is
 set to be unity ($t$=1) in the present study.
It is well known that this Hamiltonian eq. (\ref{Model}) can be mapped on an $XXZ$ quantum  spin Hamiltonian in the limit $U\rightarrow\infty$. 
The term of the nearest-neighbor interaction $V$ corresponds to 
 the $Z$-component of the antiferromagnetic exchange coupling   and the transfer
 energy $t$  corresponds to the $X$-component of that. 
When the $Z$-component is larger than the $X$-component, the system has a
 "{\it Ising}"-like symmetry and an excitation gap exists.
For the  Hubbard  model, this corresponds to the case with $V>2t$ 
 where the exact result of the charge gap is given by \cite{Yang2}
\begin{equation}
     \Delta=4(\sinh\lambda)\sum_{n=-\infty}^{\infty}\frac{(-1)^n}{2\cosh
 n\lambda} 
\label{Delta}
\end{equation}
with 
\begin{equation}
\lambda=-\ln(V/2-\sqrt{(V/2)^2-1)}).
\label{lambda}
\end{equation}
On the other hand, in the case with "$XY$"-like symmetry ($V<2t$),
the system is metallic and  the Luttinger-liquid parameter $K_\rho$ is
 exactly obtained by $\cos(\frac{\pi}{4K_{\rho}})=-V/2$ \cite{Luther}.

In order to introduce our  approach, we briefly discuss a general argument
 for 1D-electron systems  based on the  bosonization theory
 \cite{Emery1,Solyom,Voit}.
According to this theory, the effective Hamiltonian for the 1D electron
 systems  can be generally separated into  the charge  and spin parts. 
Therefore, we turn our attention to only the charge part and do not consider the 
 spin part in this work.
In the low energy limit, the effective Hamiltonian of the charge part  is 
 given by
\begin{eqnarray} 
     H_{\rho}&=&\frac{v_{\rho}}{2\pi}\int_0^L {\rm d}x
  \left[K_{\rho}(\partial_x \theta_{\rho})^2
       +K_{\rho}^{-1}(\partial_x \phi_{\rho})^2\right]   \nonumber
\\
   &+& \frac{2 g_{3\perp}}{(2\pi\alpha)^2}
  \int_0^L {\rm d}x \cos[2\sqrt{8}\phi_{\rho}(x)],
 \label{Hrho}
\end{eqnarray}
where $v_{\rho}$ and $K_{\rho}$ are the charge velocity and the coupling 
 parameter, respectively. 
The  operator $\phi_{\rho}$ and the dual operator $\theta_{\rho}$ represent
 the phase fields of the charge part. 
 $g_{3\perp}$   denotes the amplitude of the umklapp  scattering and
 $\alpha$ is a short-distance cutoff.  
On the basis of the Hamiltonian eq. (\ref{Hrho}), the electronic state is described by only the two  parameters 
 $K_{\rho}$  and $g_{3\perp}$  except for the energy scale determined by $v_{\rho}$.

At quarter filling, the $8k_F$ umklapp scattering plays the crucial effect for 
 the charge  gap.  
 The effect of the umklapp term is renormalized   under the change of
 the cutoff $\alpha\rightarrow {\rm e}^{\ell}\alpha$,
where $\ell$ is the scaling quantity.
This  process is also considered as the change of the  system size $L \to
 {\rm e}^{\ell}L$. Therefore, the size dependence of $K_{\rho}$ is described
 by the  RG equations.\cite{Sano1,Sano2,Sano3}
In this work, we adopt  the Kehrein's formulation as the RG
 equations \cite{Kehrein1,Kehrein2}
\begin{eqnarray}
 \frac{{\rm d}K_{\rho}(\ell)}{{\rm
 d}\ell}&=&-8\frac{G(\ell)^2K_{\rho}(\ell)^2}{\Gamma(8K_{\rho}(\ell)-1)},
 \label{dK} \\
 \frac{{\rm d}\log G(\ell)}{{\rm d}\ell}&=&[2-8K_{\rho}(\ell)],
  \label{dG}
\end{eqnarray}
where   $\Gamma(x)$ is $\Gamma$-function and  $G(l)$  stands the umklapp 
effect with  $G(0)=g_{3\perp}/(2\pi v_{\rho})$.
Here, the value of the short-distance cutoff $\alpha$ is selected to  a
 lattice constant of the system and set to be unity.
  This  formulation is an extension of the perturbative RG theory and 
allows us to estimate  the charge gap beyond the weak coupling regime.

To solve the  RG equations concretely, we need an initial condition for the 
 two values: $K_{\rho}(0)$ and $G(0)$.
Because it is  easy for the ED calculation to obtain $K_{\rho}(\ell)$ as
 compared to $G(\ell)$,  we  eliminate $G(\ell)$ in the RG equations. 
For this purpose, we  integrate  eq. (\ref{dG}) to yield 
\begin{equation}
G(\ell)=G(\ell_1)e^{\int_{\ell_1}^{\ell}[2-8K_{\rho}(\ell')]d\ell'}, 
\label{Gell}
\end{equation}
where  $\ell_1$ is a constant.
Substituting eq. (\ref{Gell}) into eq. (\ref{dK}), we obtain 
the differential equation for $K_{\rho}(\ell)$ as
\begin{equation}
\frac{{\rm d}K_{\rho}(\ell)}{{\rm d}\ell}=-8\frac{G^2(\ell_1)
 e^{\int_{\ell_1}^{\ell}[4-16K_{\rho}(\ell')]d\ell'}
 K_{\rho}(\ell)^2}{\Gamma(8K_{\rho}(\ell)-1)}.
 \label{Krho}
\end{equation}
Setting $K_{\rho}(\ell_1)$  as the initial  condition, we   solve 
eq.  (\ref{Krho})  numerically  except the constant $G(\ell_1)$.
The value of  $G(\ell_1)$ is  determined by comparing the solution
 $K_{\rho}(\ell)$ at $\ell =\ell_2$ with the initial value 
 $K_{\rho}(\ell_2)$.
Then, the solutions for eqs. (\ref{dK}) and (\ref{dG})  
are  completely obtained.

Using the relation $\ell \simeq \ln L$, we calculate 
two initial values $K_{\rho}(\ell_1)$ and $K_{\rho}(\ell_2)$
with  $L_1$- and $L_2$-site systems by the ED method.
In the finite size systems,  $K_{\rho}(L)$  is  calculated by  the charge
 susceptibility  $\chi_c$ and  the Drude weight $D$ by
\begin{equation}
      K\sb{\rho}=\frac{1}{2}(\pi \chi_c D)^{1/2}
\label{K}
\end{equation}
with 
$
D=\frac{\pi}{L} \frac{\partial^2 E_0(\phi)}{\partial \phi^2}
$
, where  $E_0(\phi)$ is the total energy of the ground state as a function
 of  a
 magnetic flux $L \phi$ and  $L$ is  the system size \cite{Voit}. 
Here, the magnetic flux is imposed by  introducing the following  gauge
 transformation:  $c_{m\sigma}^{\dagger} \to 
 e^{im\phi}c_{m\sigma}^{\dagger}$ for an arbitrary site $m$.
The uniform  charge  susceptibility $\chi_c$ is obtained from 
\begin{equation}
\chi_c=\frac{4/L}{E_{0}(N+2,L)+E_{0}(N-2,L)-2E_{0}(N,L)},
\label{chi}
\end{equation}
where $E_{0}(N,L)$ is the  ground state energy of a system with $L$
 sites and $N$ electrons. Here, the filling $n$ is defined  by 
 $n=N/L$.
We numerically diagonalize the Hamiltonian eq. (\ref{Model})  up to 
16 sites system 
 using the standard Lanczos algorithm. 
In the case with $U =\infty$, we also calculate $E_{0}(N,L)$ by using 
the Bethe ansatz
 method\cite{Yang1}  for finite size systems up to 800 sites system. 
Using the definitions of eqs. (\ref{K}) and  (\ref{chi}), 
we calculate  $K_{\rho}$ and $\chi_c$
 from the ground state energy of the  finite size system.


\begin{figure}[t]
  \begin{center}
\epsfxsize=6.5cm
\epsffile{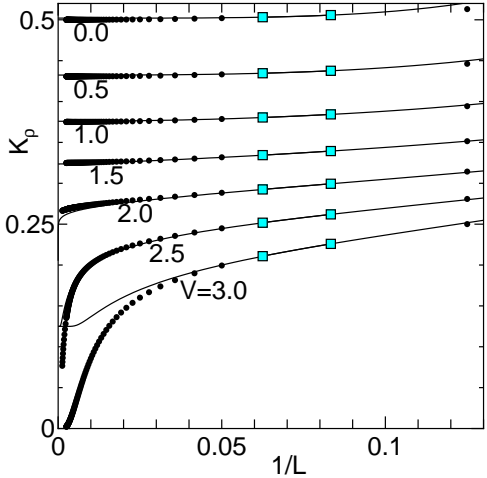}
\end{center}
  \caption[]{
   The size dependence of  $K_{\rho}(L)$ obtained from the RG equations 
 (solid lines)  and the exact Bethe ansatz results  (filled circles) for
 various $V$ at $U=\infty$.
 The shadowed squares are the numerical initial conditions with $L_1=12$ and
 $L_2=16$ for the RG equations. 
}
  \label{fig:1}
\end{figure}

In Fig. 1, we   show the size dependence of  $K_{\rho}(L)$ obtained from the
 solutions of the RG equations  together with the exact Bethe ansatz results
   for various $V$ at $U=\infty$.
We choose  $L_1=12$ and $L_2=16$  for the numerical initial condition in the
 RG equations, and we  set  $L_1=L_2-4$ hereafter.
 The limit $K_{\rho}(L \to \infty)$ of the BA result  becomes a finite value
 for $V\le 2$ and converges to  zero for $V>2$.
We see that  the RG solution is very close to the exact result and the size
 dependence of $K_{\rho}(L)$ is well  described by the RG equations   except
   very large size systems for $V>2$.

\begin{figure}[t]
  \begin{center}
\epsfxsize=6.9cm
\epsffile{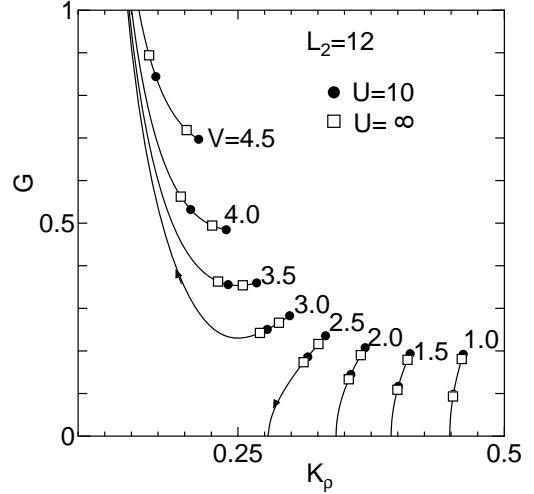}
\end{center}
  \caption[]{
   The RG flow on the $K_{\rho}-G$ plane  for various $V$ at $U=10$ with the
 numerical initial condition (filled  circles).
 The open squares are the numerical initial condition at  $U=\infty$ fitting
 to   the RG flow for $U=10$.
  }
\label{fig:2}
\end{figure}

In Fig. 2, we show  the RG flows of the systems with various $V$ for $U=10$ 
 together with those for $U=\infty$. 
 We observe that the RG flows for $U=10$ and those for $U=\infty$ 
 coincide to each other. 
Based on the Luttinger liquid theory, it is considered that the systems 
 assigned by the same RG flows have the same electronic state except energy
 scale $v_\rho$.
If we find out a  RG flow  for $U=\infty$ (we call it a reference system) 
 corresponding to that for finite $U$ (it is a original system), we can connect
 both systems through the RG flow  and derive  properties  of the original
 system from the known result of the reference system.
To identify  the RG flow, we  use  a  renormalized coupling constant
 $\tilde{G}(\ell)$  constructed by the  product of $G(\ell)$ and the
 effective energy scale  $e^{-\ell(2-8K_{\rho})}$. 
In the limit $\ell \to \infty$, $G(\ell)$ diverges in proportional to
 $e^{\ell(2-8K_{\rho})}$ (see eq. (\ref{Gell})), while $\tilde{G}(\infty)$  
 remains a  finite value  and  it becomes a unique index characterizing  the
 RG flow.

Here, we  determine the nearest neighbor repulsion $V$ in the reference system 
with $U=\infty$ so as to fit the RG flow of the reference system to that of the 
original system with original's $V$ for a finite value of $U$.
The reference's parameters $V$ corresponding to the several original's 
parameters $V$ for $U=10$ are shown in Table \ref{table1}. 
The  reference's   $V$  is smaller than the corresponding original's $V$. 
This suggests that  the  on-site repulsion $U$ causes the renormalization
 of the nearest neighbor repulsion $V$.
Fig.2 also shows that  the point  indicating  the initial condition of the
 reference system   is located at downstream than that of the original
 system on the  RG flow.  This means that the effective size  of the reference
 system  is larger than  the original system size.

\begin{table}[b]
\caption{The  original's and  the reference's parameters $V$ 
for $U=10$ (see in the text).}
\label{table1}
\begin{center}
\begin{tabular}{lcccccccc} 
Original's  $V$ & 1.0 & 1.5 & 2.0 & 2.5 & 3.0 & 3.5  & 4.0   \\  
\hline
 
Reference's $V$ & 0.38 & 0.85 &  1.33 &  1.84 &  2.38 &  2.92 &  3.43 \\ 
\end{tabular}
\end{center}
\end{table}

Substituting the reference's $V$ into eq. (\ref{lambda}), we obtain 
the charge gap $\Delta_{\rm r}$ of the reference system from eq. (\ref{Delta}). 
Taking into account the  difference of the energy scale, {\it i.e.}, 
the charge velocity between the original and the  reference systems, 
we  estimate the charge gap of the original system as
$\Delta=v_{\rho}^{\rm o}/v_{\rho}^{\rm r} \Delta_r$, where $v_{\rho}^{\rm o}$ and $v_{\rho}^{\rm r}$
 are the charge velocities of the original and the reference systems,
 respectively. 
Both of the charge velocities are calculated from the ED results with same size 
 systems through the relation of $v_{\rho}=D/2K_{\rho}$. 
 
\begin{figure}[t]
  \begin{center}
\epsfxsize=6.5cm
\epsffile{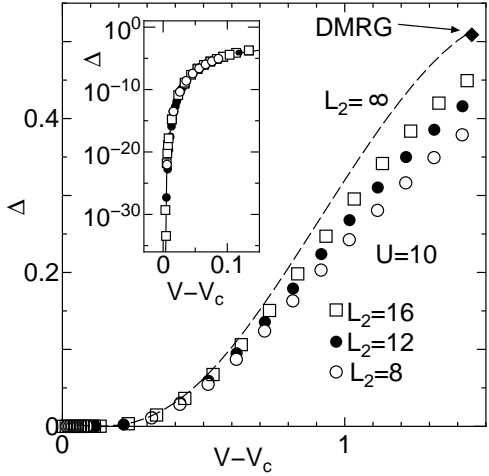}
\end{center}
    \caption[]{
The charge gap $\Delta$ as a  function of $V-V_c$  at $U=10$ for $L_2=8$, 12
 and 16  together with the extrapolated result (dashed line). 
The solid diamond is the DMRG result \cite{Ejima}.
Inset shows the semi-log plot of $\Delta$ near the MIT.
  }
\label{fig:3}
\end{figure}

In Fig. 3, we plot the charge gap $\Delta$ as a  function of $V-V_c$    at
 $U=10$ for $L_2=8$, 12  and 16  together with the  result of an extrapolation 
 with $L_2 \to \infty$.  Here,  $V_c$ is the critical value of the MIT  and
 determined by the condition  $ K_{\rho}(\ell \to \infty)=1/4$.
The values of $V_c$ are given by 2.684, 2.583 and  2.567 for finite size 
systems with $L_2=$8, 12 and 16, respectively, which yield an $L_2\to\infty$ 
extrapolated value $V_c=2.55$. 
We note  that detailed analyses of $V_c$ and the MIT have been already discussed
  in the previous works \cite{Nakamura,Sano1,Sano2,Penc,Mila,Sano0}. 
The $L_2$-dependence of $\Delta$  is assumed to be proportional to $1/L_2$, 
 resulting in  an extrapolated value of $\Delta$ with $L_2 \to \infty$ 
 as shown in Fig. 3. 
The obtained result of $\Delta$ is in good agreement  with the recent 
DMRG result of $\Delta$  at $V-V_c\simeq 1.45$ \cite{Ejima}.  
The inset in Fig. 3 shows the semi-log plot of $\Delta$ near  the MIT. 
The system size dependence of  $\Delta$ is very small even in the critical 
regime of the MIT  near  $V_c$. 
These results show that the new combined approach is especially  efficient 
to analyze the very small charge gap near the MIT.

\begin{figure}[t]
  \begin{center}
\epsfxsize=7.0cm
\epsffile{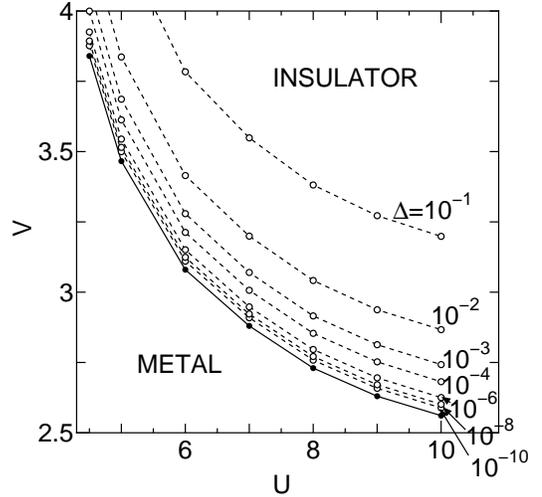}
\end{center}    
\caption[]{
The contour lines of the charge gap $\Delta$  near the MIT   on
 the $U$-$V$ plane.
The solid line  represents the phase boundary of the MIT.
}
\label{fig:4}
\end{figure}

In Fig. 4, we show the detailed results of contour lines for the extrapolated 
value of the charge  gap  $\Delta$ near  the MIT on the $U$-$V$ plane.
Rough estimation of   $\Delta$ has  been already reported  in our previous
 paper \cite{Sano3}.  
However, the previous result was limited in the case with large gap 
($\Delta>0.25$) region, since the usual finite size scaling  is used for 
the ED result.
We stress that our new combined approach has an ability to estimate the very 
small charge gap of order of $10^{-10}$  far from the limitation of the
 usual numerical estimation.


In summary, we examined the new combined approach of the ED, the RG and BA 
 methods to clarify  the  charge gap $\Delta$  of  the 1D extended Hubbard
 model with the on-site and the  nearest-neighbor interactions $U$ and $V$
 at quarter   filling. 
 Analyzing the solution of the RG equations, we connect  the original system 
 with a finite $U$  to the  reference system with $U=\infty$ in which the charge
 gap has obtained as a function of $V$ by using the exact BA method. 
Adjusting the parameter $V$ of the reference system  so as to fit the RG flow 
 of the reference system to that of the original system, we estimate the 
 absolute value of $\Delta$ of the original system including the prefactor. 
This approach is able to supply us with unambiguous and accurate result of
 $\Delta$  beyond the usual RG method and/or the ED method,  even if  energy
 scale  becomes exponentially small. 
Detailed Analysis of $\Delta$  is  shown as the contour lines  on the
 $U$-$V$ plane in the critical regime near the MIT with very small gap.

\acknowledgements
The authors would like to  thank K. Takano and T. Matsuura for useful
discussion. This work was partially supported by the Grant-in-Aid
for  Scientific Research from the Ministry of Education,
Culture, Sports, Science and Technology of Japan, 
and was performed under the interuniversity cooperative Research program 
of the Institute for Materials Research, Tohoku University.

\end{document}